\documentclass{mn2e}

\usepackage{graphicx}

\newcommand{\targ}{V407~Vul}
\newcommand{\msun}{\mathrm{M}_\odot}
\newcommand{\rsun}{\mathrm{R}_\odot}
\newcommand{\pc}{\hbox{pc}}
\newcommand{\K}{\hbox{K}}
\newcommand{\yr}{\hbox{yr}}
\newcommand{\mnras}{MNRAS}
\newcommand{\aap}{A\&A}
\newcommand{\apj}{ApJ}
\newcommand{\apjl}{ApJL}
\newcommand{\pasp}{PASP}

\title{\targ: a direct impact accretor}

\author[T. R. Marsh, D.Steeghs]
       {T. R. Marsh, D. Steeghs\\
        Department of Physics and Astronomy, University of Southampton,
Highfield, Southampton S017 1BJ}

\date{Accepted ;
      Received ;
      in original form}

\pagerange{\pageref{firstpage}--\pageref{lastpage}}
\pubyear{2001}

\begin{document}

\maketitle

\label{firstpage}

\begin{abstract}
\targ\ ($=$ RXJ1914.4+2457) shows pulsations in X-ray flux on a period
of $9.5$ minutes, which have been ascribed to accretion onto a
magnetic white dwarf, with the X-ray pulses seen as the accreting pole
moves into and out of view. The X-ray flux drops to zero between
pulses, and no other periods are seen, suggesting that \targ\ is a
type of system known as a ``polar'' in which the white dwarf has a
strong enough field to lock to the orbit of its companion.  If so,
then \targ\ has the shortest orbital period known for any binary
star. However, unlike other polars, \targ\ shows neither polarization
nor line emission. In this paper we propose that \targ\ is the first
example of a new type of X-ray emitting binary in which the mass transfer
stream directly hits a non-magnetic white dwarf as a result of the
very compact orbit. Our model naturally explains the X-ray and optical
pulsations, as well as the absence of polarization and line emission.
We show that direct impact will occur for plausible masses of the
accreting star and its companion, e.g. $M_1 \approx 0.5$, $M_2 \approx
0.1\,\msun$. In our model \targ\ retains its status as the binary star with the
shortest known orbital period, and is therefore a strong source of low-frequency
gravitational waves. \targ\ is representative of an early phase of
the evolution of the AM~CVn class of binary stars and will evolve into
the normal disc-accretion phase on a timescale of $10^6$ to
$10^7\,\yr$. The existence of \targ\ supports the double-degenerate
route for the formation of AM~CVn stars.
\end{abstract}


\section{Introduction}
Accreting white dwarfs in binary stars fall into two groups according
to the magnetic field of the white dwarf. The flow of material close
to strongly magnetic white dwarfs is controlled by the magnetic field,
and matter is channelled onto one or both magnetic poles. In this case,
accretion energy is released as the material crashes into the white
dwarf, emitting copious X-rays and optical cyclotron emission. The
X-ray and cyclotron emission are modulated on the spin period of the
white dwarf as the accreting poles rotate into and out of our
view. Moreover, the cyclotron emission is circularly and linearly
polarized. In the non-magnetic case, by contrast, the accreting
material, having some initial angular momentum, forms a disc. The disc
material accretes onto the white dwarf via an equatorial boundary
layer where the kinetic energy of the gas is released. The resulting
radiation is not significantly modulated, and there is no polarized
cyclotron radiation. Pulsed X-ray emission from white dwarf binaries
has therefore been regarded as a secure indication that the accretor
is magnetic.

When X-ray pulsations on a period of $9.5\,\min$ were discovered from
the star \targ\ (=RX J1914.4+2457), they were immediately interpreted
in terms of an accreting, magnetic white dwarf spinning on the same
period (Motch et~al. 1996).  Similar spin periods are commonly seen
in the ``intermediate polar'' class of cataclysmic variable star in
which a relatively weakly-magnetized white dwarf spins faster than the
binary orbit. However, in these stars one also sees other periods,
related to the orbital period and the ``beat'' period between the spin
and orbital periods. In \targ, however, both X-ray and optical data
show just the one period of $9.5\,\min$. Moreover, the X-ray flux from
\targ\ drops to zero between pulses, which is difficult to account for
on an intermediate polar model. This led Cropper et~al. (1998) to
suggest, instead, that \targ\ is a ``polar'', in which the white
dwarf's magnetic field locks its spin to its companion star's
orbit. If so, then \targ\ has an orbital period of only $9.5\,\min$,
the shortest known for any binary system. Such a period implies that
the donor is a helium-rich degenerate star, making \targ\ the first
magnetic member of the AM~CVn stars. The AM~CVn stars are a select
group of eight mass-transferring binary stars (other than \targ),
which have periods ranging from $17\,\min$ (AM~CVn,
Nelemans, Steeghs \& Groot 2001) to $65\,\min$ (CE~315,
Ruiz et~al. 2001). As for \targ, their short periods imply that the
donor stars are hydrogen-deficient, and indeed no hydrogen appears in
their spectra. They are thought to form from initially detached double
white dwarf systems, from systems with helium-star donors or from mass
transfer initiated when a $\sim 1\,\msun$ donor starts to transfer mass
to a white dwarf at the end of its core hydrogen burning
(Nelemans et~al. 2001; Podsiadlowski, Han \& Rappaport 2001). Double white dwarfs that
fail to become AM~CVn stars are possible Type~Ia supernova
progenitors.

A problem with the polar model is that \targ\ shows no optical
polarization (Ramsay et~al. 2000), possible on the polar model only
if the white dwarf has either a very strong or relatively weak field
(while remaining synchronized). In an effort to explain this,
Wu et~al. (2001) proposed a model in which the spin of the magnetic
white is not synchronised with the orbit leading to dissipation of
electric currents in the donor which produces unpolarized optical
flux. However, the dissipation also leads to synchronization on a
short timescale, which makes the chance of such a configuration low.

In this paper we present an alternative model for \targ, in which the
white dwarf need not be magnetic, but the X-rays will still be
strongly modulated. \targ\ may thus be the first example of a new
class of X-ray emitting binary star. We start by describing the
observational characteristics of \targ\ that need explaining.

\section{Observed features of \targ}
\label{sec:observations}
While there have been relatively few observations of \targ, any model of
the system must satisfy the following constraints:
\begin{enumerate}
\item The $9.5\,\min$ X-ray pulsations. The X-ray bright phase occupies
half of the pulsation period; for the other half of the cycle, the X-ray
flux is undetectable (Cropper et~al. 1998). No other periodic signals
are seen in X-rays.

\item The $9.5\,\min$ optical pulsations. Again, no other periodic
signals are seen. The optical pulsations have a peak-to-peak amplitude
of $0.07$ magnitudes. The optical flux peaks 0.4 cycles before the 
X-ray flux (Ramsay et~al. 2000).

\item The X-ray spectrum. This is soft and can be fitted with an absorbed
black-body spectrum of temperature $40$--$55\,\mathrm{eV}$ 
(Motch et~al. 1996; Wu et~al. 2001).

\item The lack of optical polarization. Ramsay et~al. (2000) measured
$0.3$\%\ circular polarization, a level consistent with zero given the
systematic uncertainties.

\item The optical spectrum. Unfortunately this has not yet been published,
but it is reported to be devoid of emission lines
(Wu et~al. 2001). Without having seen it, it is hard to judge the
importance of this, but we will consider it as another possible constraint
to satisfy.

\item The distance. \targ\ is heavily absorbed from which
Ramsay et~al. (2000) obtain $d > 100\,\pc$. They further find that
$d < 400\,\pc$ from a de-reddened $I$-band magnitude of $15.5$,
although the ``reddening'' is deduced from the X-ray column, and may
be too high since the colours end up being too blue even for a
Rayleigh-Jeans spectrum. Ramsay et~al. (2000) used $A_V  =5.6$, but
also state that colours deduced from $A_V = 4$ do fit black-bodies,
from which we estimate that the de-reddened $I$-magnitude lies in the
range $15.5 < I < 16.1$. 
\end{enumerate}

It was the singly periodic signal and the on/off nature of the X-ray
light-curve that led Cropper et~al. (1998) to suggest their polar
model. The absence of both polarization and optical line emission
presented difficulties that led Wu et~al. (2001) to develop their
unipolar inductor model by analogy with the Jupiter-Io system. As
outlined in the introduction, in their model, the magnetic white dwarf
(the accretor in our model) is slightly asynchronous with the binary
orbit and the resulting electric field drives currents that run
between the two stars, leading to energy dissipation in both of
them. Dissipation on the magnetic white dwarf powers the X-ray
emission, while dissipation at the donor plus irradiation powers the
optical flux. The irradiation is predicted to be comparable to the
ohmic dissipation, which they suggest can explain the relatively weak
optical modulation and the lack of line emission.  Since the
non-magnetic star dominates the optical emission, the absence of
polarization is also explained.

The major problem with Wu et~al.'s model is that it is
short-lived: they estimate that it will last only $\sim
1000\,\yr$. Even if all AM~CVn systems pass through this
stage, there would only be 1--10 such systems in our Galaxy
at any one time, according to the formation rates of
Nelemans et~al. (2001) and Podsiadlowski et~al. (2001). The chance of
finding one within our neighbourhood is therefore small. Magnetic 
systems may in fact comprise only a small fraction of
the total, making the probability of finding such a system very low.
It is therefore worth searching for longer-lived models.

\section{A new model for \targ}
\label{sec:model}
Our idea is simple: in very close binary systems, the mass transfer
stream can plough straight into the accretor even in the absence of
the magnetic field. This happens if the minimum distance of the
ballistic gas stream from the centre of mass of the accretor is
smaller than the accretor's radius, which is most famously the case in
Algol binary stars, where the accretors are main-sequence stars. We
propose that \targ\ is the first instance of Algol-like direct impact
in the case of a white dwarf accretor.

\begin{figure}
\includegraphics[width=\columnwidth]{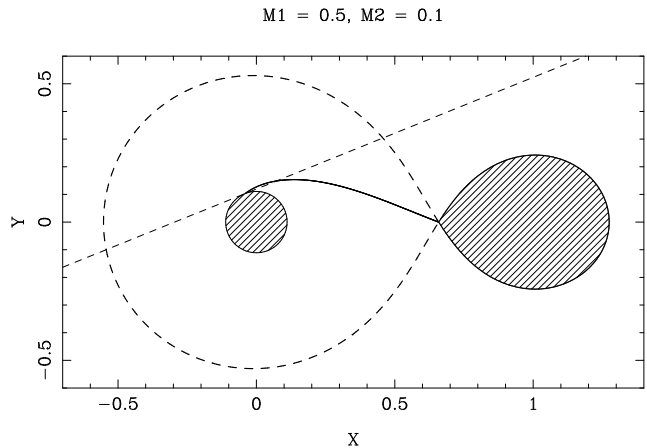}
\caption{Path of the stream in \targ\ in the case $M_1 = 0.5\,\msun$, 
$M_2 = 0.1\,\msun$. The dashed line is tangent at the impact point, 
to show that the impact is hidden from the donor in this case.
 \label{fig:binary}}
\end{figure}
In Fig.~\ref{fig:binary} we show that the direct impact is possible
using typical system parameters (section~\ref{sec:param}). The figure
shows the ballistic path of the mass transfer stream calculated for an
accretor mass of $M_1 = 0.5\,\msun$ and a donor mass of $M_2 =
0.1\,\msun$. The separation of the binary is fixed through Kepler's
laws by the masses and the orbital period, $P = 9.5\,\min$, while we
have used Nauenberg's (1972)
analytic formula for the radius of the white dwarf. If the $9.5\,\min$
pulsations truly reflect \targ's orbital period, direct stream impact
is possible.

This model immediately explains the absence of polarization and the
X-ray pulsations. There is no polarization because the white dwarf is
non-magnetic, while the impact point is fixed in the rotating frame of
the binary and will produce one pulse per orbit. Our model nicely
explains the absence of X-ray flux for half the cycle since the impact
will be on the equator of the accretor; this is a key problem with
intermediate polar models of \targ, as Cropper et~al. (1998) point
out.

There are two possible explanations for the optical pulsations.
First, we do not expect the white dwarf to be synchronized with the
orbit, but instead to spin rapidly as the result of the accretion of
material of high specific angular momentum.  The surface of the white
dwarf will therefore move under the point of impact, and one expects a
heated area to trail downstream from the spot. If this area is
responsible for the optical pulsations, they would then peak at an
earlier phase than the X-rays, as observed. The amplitude of the
optical pulsations in this case depends upon the extent of the heated
region and the brightness of the unheated photosphere of the white
dwarf. In order to dominate the optical flux, given the lower limit
on the distance of $100\,\pc$ (Ramsay et~al. 2000) and the unreddened
$I$ magnitude between $15.5$ and $16.1$
(section~\ref{sec:observations}), we require $M_I < 10.5$ -- $11.1$.
For a $0.5\,\msun$ white dwarf this implies an effective temperature
$T_\mathrm{eff} > 15$,$000$ -- $21$,$000\,\K$ (Bergeron, Wesemael \& Beauchamp 1995),
which is nothing out of the ordinary. Indeed, given an accretion rate
of $10^{-8}\,\msun\,\yr^{-1}$ (section~\ref{sec:evln}), we would
expect a much higher temperature from compressional heating alone:
Townsley \& Bildsten (2002) give a surface temperature of $27$,$000\,\K$
from compressional heating for a $0.6\,\msun$ white dwarf accreting at
$10^{-9}\,\msun\,\yr^{-1}$.

Alternatively, the optical pulsations may come from the heated face of
the donor star. This could be heated by irradiation from the impact
spot (but see the next section), or by photospheric emission from the
white dwarf. Although the latter is normally ignored, once again the
very close orbit of \targ\ means that it is not just possible, but
likely. The geometry of the impact site means that the heated face of
the donor star will also produce a peak optical flux in advance of the
X-ray flux, as observed, although the impact site would have to be on
the side facing the donor to explain the $0.4$ cycle shift seen. The
temperature of the heated face of the donor due only to photospheric
emission from the accretor is $T_\mathrm{h} \approx (R_1/a)^{1/2} T_1
\approx 0.33 T_1$, where $R_1$ and $T_1$ are the radius and
temperature of the accretor. Given that the donor is about 3 times the
size of the accretor, and the temperature of the accretor estimated
above, the donor could easily produce the optical pulsations.

Of the observational facts above, we are left to explain the reported
lack of optical line emission and the softness of the X-ray spectrum.

\subsection{Optical Line Emission}
The direct-impact model provides a beautiful way to avoid X-ray
irradiation of the donor star, and therefore any associated emission
lines, altogether, because it is possible for the impact point to be
hidden from the donor star.  Fig.~\ref{fig:binary} illustrates just
such a case; we will examine the parameter constraints needed for this
to be the case in the next section. Nevertheless, it is not clear
whether it is really necessary to exclude irradiation of the donor
because any emission lines would execute high-amplitude sinusoidal
motion on a period of only $9.5\,\min$ and would be smeared in
wavelength, especially given that \targ\ is optically faint (the
orbital velocity of the donor in Fig.~\ref{fig:binary} is
$800\,\mathrm{km}\,\mathrm{s}^{-1}$).  Given the hydrogen deficiency
and the proximity of the two stars, one would only expect to see
ionized helium emission. Since \targ\ is only easily observable in the
$I$-band, it is not clear whether any emission lines should have been
detected, even if irradiation is taking place. There is, of course,
some irradiation of the stream, but only where it nears the impact
site. By this time it is moving very fast, and will be highly ionized,
and since emission from this part is hard to detect even in much brighter
and longer period polars, we do not regard this as a problem.

\subsection{The Soft X-ray Spectrum}
Direct impact accretion differs significantly from
magnetically-confined accretion. First, the accretion stream will be
much narrower in the direct impact case. The reason is that in the
magnetic case, threading occurs over a range of radii leading to
impact in a relatively large arc close to one or both poles of the
white dwarf.  On the other hand, as Lubow \& Shu (1976) and
Lubow (1989) showed, as the ballistic stream nears its
closest approach to the centre of mass of the accretor, it
becomes comparable in width to the hydrostatic scale height that an accretion
disc would have at the same radius. At the white dwarf this is a width
of order $\sim (kT R_1^3/GM_1 m_p)^{1/2} \approx 5 \times 10^{-3}\,R_1$ for
$T = 20$,$000\,\K$ (stream from the heated face of donor) and $M_1 =
0.5\,\msun$. Therefore, accretion will take place over a fraction 
$f \sim 0.01$\%\ of the white dwarf's surface.

Together with an accretion rate of $\dot{M} \sim
10^{-8}\,\msun\,\yr^{-1}$ (section~\ref{sec:evln}), some 10--100
times higher than in polars, the narrow stream width makes the
accretion in \targ\ comparable to the blob accretion model
(Kuijpers \& Pringle 1982;  Frank, King \& Lasota 1988;  King 2000), in which the soft X-ray
excess of polars is explained by the stream breaking into dense blobs
which are able to penetrate below the photosphere and therefore become
thermalized, giving soft X-ray emission. This happens for blob
densities $\rho > 10^{-7}\,\mathrm{g}\,\mathrm{cm}^{-3}$
(Frank et~al. 1988). For a stream width of $10^{-4}\,\rsun$, an
accretion rate of $\dot{M} = 10^{-8}\,\msun\,\yr^{-1}$, and a
free-fall velocity $v_{ff} = 4 \times
10^8\,\mathrm{cm}\,\mathrm{s}^{-1}$, we find $\rho \approx 3 \times
10^{-5} \,\mathrm{g}\,\mathrm{cm}^{-3}$, well in excess of
Frank et~al.'s limit.

In magnetically-confined accretion, after the accretion shock, the material
continues to flow along the field lines and can only come to a halt
through cyclotron and bremsstrahlung cooling. This leads to the shock
forming some height above the point at which the ram pressure of the
stream matches the pressure in the white dwarf's atmosphere. In the
direct impact case, there is nothing to stop the stream expanding
sideways following the shock which will cause adiabatic cooling and
lead to the shock forming much closer to the point where the pressures
match. This increases the chance that the shock will be buried.  The
final, and perhaps most important, difference is that the white dwarf
moves rapidly beneath the impact site, which will sweep the heated
material downstream. We suggest that it is this which spreads the
emission over a large enough area to cause the spectrum to be very soft as
observed. We can crudely estimate the cooling time as the thermal
timescale of a column of gas heated to the observed temperature of $T
\approx 50\,\mathrm{eV}$. The column density $\Sigma$ is fixed by the
penetration depth according to
\begin{equation}
\Sigma g = P_\mathrm{ram} = \rho v_\mathrm{ff}^2,
\end{equation}
where $g = GM_1/R_1^2$ is the surface gravity, $\rho$ is the stream density and
$v_\mathrm{ff}$ is the free-fall velocity ($v_\mathrm{ff}^2 =
2GM_1/R_1$), which is comparable to that of the stream. Given an
energy of $3kT/2$ per particle, and a loss rate of $\sigma T^4$, we
obtain a cooling time $t_c$, in seconds, of
\begin{equation}
t_c \approx 1.0
\left(\frac{\rho}{3\times 10^{-5}\,\mathrm{g}\,\mathrm{cm}^{-3}}\right)
\left(\frac{R_1}{0.01\,\rsun}\right)
\left(\frac{T}{50\,\mathrm{eV}}\right)^{-3} 
\end{equation}
Since the white dwarf spin period may be as short as 10 seconds, this
cooling time is the right order of magnitude to lead to significant
spreading of the emission. All that is needed is that the area is
spread so that $f \sim 0.1$\%, for the energy liberated
by accretion at the rate $10^{-8}\,\msun\,\yr^{-1}$ to be
radiated away with a temperature of $T \sim 50\,\mathrm{eV}$, as observed.

\section{Parameter Constraints}
\label{sec:param}
We now calculate the limits upon the parameters required for our model
to be viable, fixing the orbital period at $P = 9.5\,\min$. We have first
the orbital separation in terms of the total mass of the binary
\begin{equation}
a = 0.148 \left( \frac{M_1+M_2}{\msun}\right)^{1/3} \; \rsun .
\end{equation}
For the radii of the white dwarfs we use Nauenberg's
1972 formula
\begin{equation}
\frac{R}{\rsun} = 0.0112 \left[ 
\left( \frac{M}{1.433 \,\msun} \right)^{-2/3} -
\left( \frac{M}{1.433 \,\msun} \right)^{2/3} \right]^{1/2} .
\end{equation}
This is a lower limit, since Nauenberg's relation is based upon
zero temperature and no rotation, and departures from these 
assumptions act to increase the radius for a given mass.
The radius of closest approach to the centre of the white
dwarf relative to the separation of the binary is a function of 
mass ratio, for which Nelemans et~al. (2001) give a fit
based upon the calculations of Lubow \& Shu (1975), and confirmed
by our own integrations.

Next, we have the well-known relation between the mean density of the
Roche lobe filling donor and the orbital period, which gives
$\bar{\rho} = 4.3 \times 10^3\,\mathrm{g}\,\mathrm{cm}^{-3}$ in the
case of \targ.  Using Nauenberg's formula, this corresponds to
a mass of $M_2 = 0.077\,\msun$. More generally this is a lower limit
because any hydrogen content or semi-degeneracy reduces the density
for a given mass, and since the density increases monotonically with
mass, the mass must increase to match a fixed density.

Two other constraints come from considerations of the stability of
mass transfer which set upper limits to the mass ratio. To ensure that
the Roche lobe does not shrink faster than the donor in the case of
conservative mass transfer, the mass ratio must satisfy
\[ q < \frac{5}{6} + \frac{1}{2} \frac{d\ln R_2}{d \ln M_2},\]
e.g. Nelemans et~al. (2001). Nauenberg's formula gives
\[ \frac{d\ln R_2}{d \ln M_2} = - \frac{1}{3} \left( \frac{1 + x^{4/3}}{1 -
x^{4/3}}\right),\] where $x = M_1/1.433$. This strictly only applies
to the limiting case of $M_2 = 0.077\,\msun$, but we will take it to
apply to higher mass cases too. This is a conservative assumption,
because radiative stars shrink on mass loss. A stricter constraint
applies if accretion occurs directly onto the white dwarf and the
angular momentum of the stream is \emph{not}
transferred back from the white dwarf to the orbit (i.e.\ tidal
and magnetic coupling are ineffective). The limit then becomes
\begin{equation}
q < \frac{5}{6} + \frac{1}{2} \frac{d\ln R_2}{d \ln M_2}
- \sqrt{(1+q) r_h} , \label{eq:rig}
\end{equation}
(Nelemans et~al. 2001), where $r_h$ is the circularization radius as 
a fraction of $a$, which is a function of mass ratio for which we use
the approximation of Verbunt \& Rappaport (1988), accounting for their inverted definition
of $q$.

Finally, if the reported absence of emission lines is significant (see
section~\ref{sec:model} for a discussion of this), there is another
upper limit on the mass of the donor for a given accretor mass,
required in order for the impact spot to be hidden from the donor
star.  We carried out our own integrations in the Roche potential to
derive this upper limit as a function of accretor mass.
 
\begin{figure}
\includegraphics[width=\columnwidth]{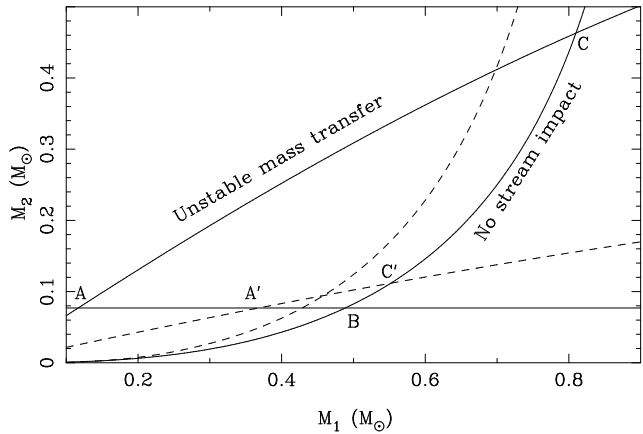}
\caption{Parameter space constraints. The almost-straight solid
and dashed lines are upper limits from mass transfer stability. 
The dashed line applies if the angular momentum of the
accreting material is not fed back to the orbit. The horizontal line
is the lower limit on the donor mass. The curved solid line is a lower
limit required for direct stream impact. The curved dashed line is the
upper limit for the impact site to be out of sight looking from the
donor star. \label{fig:allowed}}
\end{figure}
These constraints give the allowed parameter space indicated in
Fig.~\ref{fig:allowed} which shows that direct stream impact in a stable
binary is possible if the masses of the two stars lie in either the
large triangle ABC for the standard stability limit, or in the smaller
triangle A'BC' if the more rigorous stability criterion,
Eq.~\ref{eq:rig}, applies.
This smaller region is also consistent with the requirement
that the donor not be irradiated (curved dashed line in
Fig.~\ref{fig:allowed}). Moreover, this happens for $M_1 \approx
0.5\,\msun$, a typical mass for a white dwarf.  Thus this model is
feasible without the need for fine tuning of the parameters.  If the
donor is degenerate, then, for no irradiation, the parameter space 
is restricted to $M_2 = 0.077\,\msun$, $M_1 = 0.42$ -- $0.49\,\msun$.

\section{Future Evolution}
\label{sec:evln}
\targ's destiny is to become a standard member of the AM~CVn systems.
As the donor's mass decreases, even if the accretor fails to increase
in mass, the increase in the orbital period and the decrease in mass
ratio, both act to make it less likely that the stream will hit the
accretor directly.  Eventually, the stream will orbit the accretor and
a disc will form. Depending upon the precise parameters, it will do so
after the loss of $0.01$ -- $0.1\,\msun$ from the donor. The mass
transfer rate for an orbital period of $9.5\,\min$ is $\approx
10^{-8}\,\msun\,\yr^{-1}$ (Nelemans et~al. 2001), so this phase will
last $10^6$ -- $10^7\,\yr$. If all AM~CVn systems pass through such a
phase, then, according to the formation rates of Nelemans et~al. (2001)
and Podsiadlowski et~al. (2001), there should be $10^3$--$10^5$
systems like \targ\ in the Galaxy. The nearest of these should be
between $100$ and $400\,\pc$ away, consistent with estimates of the
distance to \targ. This agreement favours the detatched double white
dwarf route for the formation of AM~CVn's over either the helium
star or the semi-degenerate routes, since the latter rarely reach
orbital periods as short as 10 minutes
(Nelemans et~al. 2001; Podsiadlowski et~al. 2001), whereas
Nelemans et~al. (2001) predict that most double white dwarfs will pass
through a direct-impact phase if they avoid merging.

\section{Gravitational Waves}
\targ\ is a very promising source of gravitational waves.  Following
Meliani, de Araujo \& Aguiar (2000), who calculate strain amplitudes for many
cataclysmic variable stars, and using $M_1 = 0.5\,\msun$ and $M_2 =
0.1\,\msun$, \targ\ will produce a strain amplitude $h = 1.2 \times
10^{-21} (d/100\,\pc)^{-1}$ at Earth.  This is comparable to the
largest values listed by Meliani et~al. (2000). In addition, the very
short period of \targ\ will lead to gravitational wave emission at a
frequency of $f_\mathrm{gw} = 2/P = 3.5 \times 10^{-3}\,\mathrm{Hz}$,
where the background noise of an instrument such as LISA is predicted
to be much lower than at the lower frequencies characteristic of other
cataclysmic variable stars. \targ\ is thus one of the brightest
prospects for space-based gravitational wave detection.

\section{Conclusions}
We have shown that plausible masses for the two components of \targ\
suggest that the mass transfer stream in this system strikes the
accreting white dwarf directly. The resulting spot explains the X-ray
pulses observed from this system, without the need for a magnetic
white dwarf. This is consistent with the lack of polarization from the
system, and with the complete disappearance of X-ray flux in between pulses.
Our model has a lifetime of $10^6$ to $10^7\,\yr$ compared with the
$10^3\,\yr$ of Wu et~al.'s unipolar inductor model,
making it easier to reconcile with estimated formation rates for the
AM~CVn systems, as long as a substantial fraction of these systems
pass through a direct impact-phase, as predicted by
Nelemans et~al. (2001) on the basis of a double white dwarf origin for
these systems. \targ\ is the first example of a new class of
stream-fed, non-magnetic white dwarf accretors.

\label{lastpage}

\end{document}